\documentclass[12pt]{iopart}
\usepackage{iopams}  
\usepackage{graphicx}% Include figure files
\usepackage{color}% 

\begin{document}

\title[Dirty two-band superconductivity]{Dirty two-band superconductivity with interband pairing order}

\author{Yasuhiro Asano$^{1,2,3}$, Akihiro Sasaki$^{1}$, Alexander A.\ Golubov$^{3,4}$}

\address{$^1$ Department of Applied Physics, Hokkaido University, Sapporo 060-8628, Japan}
\address{$^2$ Center of Topological Science and Technology, Hokkaido University, Sapporo 060-8628, Japan}
\address{$^3$ Moscow Institute of Physics and Technology, 141700 Dolgoprudny, Russia}
\address{$^4$ Faculty of Science and Technology and MESA+ Institute for Nanotechnology, 
University of Twente, 7500 AE Enschede, The Netherlands}
\ead{asano@eng.hokudai.ac.jp}
\vspace{10pt}
%\begin{indented}
%\item[]August 2017
%\end{indented}

\begin{abstract}
We study theoretically the effects of random nonmagnetic impurities on the superconducting transition temperature $T_c$
in a two-band superconductor characterized by an equal-time $s$-wave interband pairing order parameter.
Because of the two-band degree of freedom, it is possible to define a spin-triplet $s$-wave pairing order 
parameter as well as a spin-singlet $s$-wave order parameter.
The former belongs to odd-band-parity symmetry class, whereas the latter belongs to even-band-parity symmetry class.
In a spin-singlet superconductor, $T_c$ is insensitive to the impurity concentration when we estimate 
the self-energy due to the random impurity potential within the Born approximation.
On the other hand in a spin-triplet superconductor, $T_c$ decreases with the increase of the 
impurity concentration. We conclude that Cooper pairs belonging to odd-band-parity symmetry class 
are fragile under the random impurity potential even though they have $s$-wave pairing symmetry.
\end{abstract}

%\maketitle

%*********************************************************************************
\section{Introduction}

%*********************************************************************************

Conventional wisdom suggests that
the dependence of superconducting transition temperature $T_c$ on the concentration 
of nonmagnetic impurities is closely related to the momentum-symmetry of the pair potential.
It is well known that $T_c$
of an $s$-wave superconductor is insensitive to the impurity concentration.
\cite{agd,abrikosov:jetp1959,anderson:jpcs1959} 
On the other hand, 
unconventional superconductivity such as $p$- and $d$-wave symmetry is fragile in the presence of impurities.
 The robustness of an $s$-wave Cooper pair under potential disorder, however, 
may be weakened in a two-band superconductor as discussed in previous 
literature~\cite{allen:review1982,golubov:prb1997,onari:prl2009,efremov:prb2011,korshunov:prb2014,
hoyer:prb2015,asano172bandimp}. 
In these papers, the \textsl{intraband} pairing order is assumed in each conduction band. 
Namely, two electrons at the first (second) band form the pair potential $\Delta_1$ 
($\Delta_2$). Such theoretical model would describe the superconducting states in
 MgB$_2$~\cite{mgb2:akimitsu2001,mgb2:louie2002} and iron pnictides~\cite{pnictide:hosono2008,hosono:physicac2015}. 
The suppression of $T_c$ by the interband impurity scatterings is a common conclusion 
of all the theoretical studies.

In addition to the \textsl{intraband} pair potentials,
the \textsl{interband} (or interorbital) Cooper pairing order has been discussed in a 
topological superconductor Cu$_x$Bi$_2$Se$_3$~\cite{hor:prl2010,fu:prl2010,sato:repprogphys2017}.
Various types of multiband superconductivity would be expected 
in topological-material based superconductors 
because the band-crossing plays an essential role in realizing the 
topologically nontrivial states.  
Moreover, a possibility of interband/interorbital Cooper pairing is pointed out also in 
a heavy fermionic superconductor UPt$_3$~\cite{upt3:rmp2002,yanase:prb2016} and 
an antiperovskite superconductor Sr$_{3-x}$SnO.\cite{oudah:nc2016}. 
%These conditions require a relation between the dispersion in the first band 
% and that in the second band.
%For instance, the dispersions $\xi_1(\bi{k})=\xi_2(\bi{-k})$ 
%fulfill these conditions. 
%The similar dispersion relations are expected 
%in a Weyl metal breaking inversion symmetry~\cite{murakami,halasz}. 
In addition to the spin-singlet order parameter, the spin-orbit coupling 
 may make the spin-triplet order parameter possible. 
Thus a superconductor with the interband pairing order can be a superconductor of a novel class.
So far, however, little attention has been paid to physical phenomena unique to an interband superconductor.

In this paper, we theoretically study the effects of nonmagnetic random impurities 
on $T_c$ in a two-band superconductor characterized by an equal-time $s$-wave interband pairing order.
The pair potential is defined by the product of two annihilation operators of 
an electron. 
Therefore, the pair potential must be antisymmetric under the commutation 
of the two annihilation operators, which is the requirement from the Fermi-Dirac statistics 
of electrons.
Due to the two-band degree of freedom, 
a spin-triplet $s$-wave pair potential is allowed as well as a spin-singlet $s$-wave one. 
The latter is symmetric under the permutation of the two band indices (even-band-parity), 
whereas 
the former is antisymmetric (odd-band-parity). 
The effects of impurity potential are considered through the self-energy 
estimated within the Born approximation. The transition temperature is calculated 
from the linearized gap equation. 
We find that $T_c$ is insensitive to the impurity concentration in a spin-singlet $s$-wave interband superconductor.
However, $T_c$ in a spin-triplet $s$-wave case decreases with the increase of the impurity concentration.
 We conclude that odd-band parity Cooper pairs 
are fragile under the potential disorder even though they belong to $s$-wave symmetry class.

This paper is organized as follows. In Sec.~2, we explain the normal state that 
makes possible spatially uniform interband Cooper pairing orders. 
The gap equation in the clean limit is 
derived for both a spin-singlet superconductor and a spin-triplet superconductor.
The effects of random impurities 
on the superconducting transition temperature are studied in Sec.~3. 
%We discuss the applicability of the obtained results in Sec.~IV. 
The conclusion is given in Sec.~4.
Throughout this paper, we use the units of $k_\mathrm{B}=c=\hbar=1$, where $k_{\mathrm{B}}$ 
is the Boltzmann constant and $c$ is the speed of light.

%==========================================================================
\section{Interband pairing order}\label{model}
%==========================================================================
%---------------------------------------------------------------------------
The interband $s$-wave pair potential is defined by
\begin{eqnarray}
\Delta_{1,\sigma; 2, \sigma'}(\bi{r}) = g \left\langle \psi_{1,\sigma}(\bi{r}) \psi_{2,\sigma'}(\bi{r}) \right\rangle,
\end{eqnarray}
where $\psi_{\lambda,\sigma}^\dagger(\bi{r})$ ($\psi_{\lambda,\sigma}(\bi{r})$) is the  
creation (annihilation) operator of an electron with spin $\sigma$ ($=\uparrow$ or $\downarrow$) at the $\lambda$ th 
conduction band and $g>0$ represents the interband attractive interaction.
By applying the Fourier transformation,
\begin{eqnarray}
\psi_{\lambda,\sigma}(\bi{r}) =\frac{1}{\sqrt{V_\mathrm{vol}}} \sum_{\bi{k}} 
\psi_{\lambda,\sigma}(\bi{k})
e^{i\bi{k}\cdot \bi{r}},
\end{eqnarray}
the pair potential becomes
\begin{eqnarray}
\Delta_{1,\sigma; 2, \sigma'}(\bi{r}) &=&  \frac{g}{V_\mathrm{vol}} \sum_{\bi{k}, \bi{k}'} 
\left\langle \psi_{1,\sigma}(\bi{k}) \psi_{2,\sigma'}(\bi{k}') \right\rangle
e^{i(\bi{k}+\bi{k}')\cdot \bi{r}},\\
 &=& \frac{g}{V_\mathrm{vol}} \sum_{\bi{k}} 
\left\langle \psi_{1,\sigma}(\bi{k}) \psi_{2,\sigma'}(-\bi{k}) \right\rangle.\label{delta1}
\end{eqnarray}
In the second line, we assume 
the spatially uniform order parameter which is realized at $\bi{k}+\bi{k}'=0$.
To apply the weak coupling mean-field theory,
the state at $\bi{k}$ with spin $\sigma$ in the first band and 
the state at $-\bi{k}$ with spin $\sigma'$ in the second band 
must be degenerate at the Fermi level. 
Otherwise interband Cooper pairs have the center-of-mass momenta and their order parameter 
oscillates in real space~\cite{caldas:prb2012,fulde_fflo,larkin_fflo}.
Thus the interband pair potential requires a characteristic band structure. 
In this paper, we consider a normal state described by the Hamiltonian,
\begin{eqnarray}
\check{\mathcal{H}}_{\mathrm{N}} &=& 
\int d\bi{r}
\left[\psi_{1,\uparrow}^\dagger(\bi{r}), \psi_{1,\downarrow}^\dagger(\bi{r}), 
\psi_{2,\uparrow}^\dagger(\bi{r}), \psi_{2,\downarrow}^\dagger(\bi{r})\right] 
%\nonumber\\
%&\times&
\check{H}_{\mathrm{N}}(\bi{r})
\left[\begin{array}{c}
\psi_{1,\uparrow}(\bi{r})\\
\psi_{1,\downarrow}(\bi{r})\\
\psi_{2,\uparrow}(\bi{r})\\
\psi_{2,\downarrow}(\bi{r})
\end{array}\right], \label{hn_operator}\\
\check{H}_{\mathrm{N}}(\bi{r})&=&\left[ \begin{array}{cc}
\xi(\bi{r})\, \hat{\sigma}_0 & v\, e^{i\theta} \, \hat{\sigma}_0 \\
v \, e^{-i\theta}\, \hat{\sigma}_0 & \xi(\bi{r}) \hat{\sigma}_0
\end{array} \right], \label{hn}\\
\xi(\bi{r})&=& -\frac{\nabla^2}{2m}- \mu,\label{xi1}
%\xi_2(\bi{k})=& \frac{1}{2m}\left[ (\bi{k}+\bi{k}_0)^2 - k_F^2 \right].\label{xi2}
\end{eqnarray}
where $m$ is the mass of an electron, $\mu$ is the chemical potential, and 
$v$ represents the hybridization between
the two conduction bands.
Generally speaking, the hybridization potential is a complex number characterized by a phase 
$\theta$. We will show that observable values in a superconductor are independent of $\theta$ 
although the expression of the Green function depends on it.
Throughout this paper, Pauli matrices in spin, two-band, particle-hole spaces are 
denoted by $\hat{\sigma}_j$, $\hat{\rho}_j$, and $\hat{\tau}_j$ for $j=1-3$, respectively. 
In addition, $\hat{\sigma}_0$, $\hat{\rho}_0$, and $\hat{\tau}_0$ are the unit matrices 
in these spaces.
Since the two bands are identical to each other, the Hamiltonian preserves the symmetry 
described by
\begin{eqnarray} 
& \check{\Gamma}\, \check{H}_{\mathrm{N}}(\bi{r})\, \check{\Gamma}^{-1}
=\check{H}_{\mathrm{N}}(\bi{r}), \label{gamma_symmetry}\\ 
&\check{\Gamma} = \mathcal{T} \, \hat{\rho}_1, \quad 
\mathcal{T} = i\, \hat{\sigma}_2\,  \mathcal{K},
\end{eqnarray}
where $\mathcal{T}$ is the time-reversal operator, $\mathcal{K}$ means the complex conjugation. 
Thus $\Gamma$ represents the combined operation of the time-reversal and
the exchange between the two bands. 
The normal state Hamiltonian in Eq.~(\ref{hn}) is simplest model which satisfies 
Eq.~(\ref{gamma_symmetry}). The conclusions of this paper are insensitive to the normal state 
Hamiltonian. We will explain the reasons after reaching the main results. 
The electronic structure given in Eq.~(\ref{hn}) may poses both the interband and the intraband $s$-wave order parameters in its superconducting phase.
The effects of potential disorder on $T_c$ for intraband superconductivity have been already 
studied theoretically in previous papers~\cite{golubov:prb1997,onari:prl2009,efremov:prb2011,korshunov:prb2014,
hoyer:prb2015,asano172bandimp}. In our model, the amplitudes of two intraband pair potentials are expected be 
equal to each other because of the symmetry in the two conduction bands. 
It has been well established that $T_c$ of intraband superconductivity in such symmetric case 
is insensitive to the impurity scatterings.~\cite{golubov:prb1997,efremov:prb2011,asano172bandimp} 
Thus we focus only on interband superconductivity in this paper.
%The argument below is valid as far as the normal state Hamiltonian satisfies in Eq.~(\ref{gamma_symmetry}).

According to Eq.~(\ref{delta1}), we define the spatially uniform superconducting order parameter explicitly as
\begin{eqnarray}
\Delta\equiv &  \frac{g}{V_{\mathrm{vol}}} \sum_{\bi{k}}
\left\langle \psi_{1, \uparrow}(\bi{k}) \, \psi_{2, \downarrow}(-\bi{k}) \right\rangle.\label{del_def}
\end{eqnarray}
In the two-band model, it is possible to define two types of interband pairing order: spin-singlet and spin-triplet. 
In spin-singlet symmetry, the pair potential in Eq.~(\ref{del_def}) is 
symmetric (antisymmetric) under the permutation of band (spin) indices 
\begin{eqnarray}
\Delta &=& - \frac{g}{V_{\mathrm{vol}}} \sum_{\bi{k}}
\left\langle \psi_{1, \downarrow}(\bi{k}) \, \psi_{2, \uparrow}(-\bi{k}) \right\rangle
%,\\
%&=&
=\frac{g}{V_{\mathrm{vol}}} \sum_{\bi{k}}
\left\langle \psi_{2, \uparrow}(\bi{k}) \, \psi_{1, \downarrow}(-\bi{k}) \right\rangle.
\end{eqnarray}
On the other hand in spin-triplet symmetry, the pair potential in Eq.~(\ref{del_def}) is 
antisymmetric (symmetric) under the permutation of band (spin) indices 
\begin{eqnarray}
\Delta&=& \frac{g}{V_{\mathrm{vol}}} \sum_{\bi{k}}
\left\langle \psi_{1, \downarrow}(\bi{k}) \, \psi_{2, \uparrow}(-\bi{k}) \right\rangle
%,\\
%&=&
= -\frac{g}{V_{\mathrm{vol}}} \sum_{\bi{k}}
\left\langle \psi_{2, \uparrow}(\bi{k}) \, \psi_{1, \downarrow}(-\bi{k}) \right\rangle.
\end{eqnarray}
In what follows, we consider opposite-spin-triplet pairing order. 
%This, however, does not loose any generality.
%The physics discussed below can be applied also to a superconductor with equal-spin-triplet order parameter
%given by
%\begin{eqnarray}
%\Delta_{\mathrm{esp}}\equiv & \frac{g}{V_{\mathrm{vol}}} \sum_{\bi{k}}
%\left\langle \psi_{1, \sigma}(\bi{k}) \, \psi_{2, \sigma}(-\bi{k}) \right\rangle,\\
%=&-\frac{g}{V_{\mathrm{vol}}} \sum_{\bi{k}}
%\left\langle \psi_{2, \sigma}(\bi{k}) \, \psi_{1, \sigma}(-\bi{k}) \right\rangle,
%\end{eqnarray}
%with $\sigma=\uparrow$ or $\downarrow$. 
The Bogoliubov-de Gennes (BdG) Hamiltonian in momentum space is represented by 
\begin{eqnarray}
\bar{H}_{\mathrm{S(T)}}(\bi{k}) &=&\left[ \begin{array}{cc} \check{H}_N(\bi{k}) & 
\check{\Delta}_{\mathrm{S(T)}} \\ -\check{\Delta}_{\mathrm{S(T)}} & -\check{H}_N^\ast(-\bi{k})
\end{array}\right],\\
\check{\Delta}_{\mathrm{S}} & = & \Delta\, \hat{\rho}_1\, i\, \hat{\sigma}_2, \quad
\check{\Delta}_{\mathrm{T}}= \Delta\, i \, \hat{\rho}_2 \, \hat{\sigma}_1,
\end{eqnarray}
where $\check{\Delta}_{\mathrm{S}}$ and $\check{\Delta}_{\mathrm{T}}$ represent the 
spin-singlet pair potential and the spin-triplet one, respectively.
Hereafter we fix the superconducting phase at zero for simplicity. 
%It is easy to confirm the symmetry relation,
%\begin{eqnarray}
%&\bar{\Gamma} \, \bar{H}_{\mathrm{S(T)}}(\bi{k})\,  \bar{\Gamma}^{-1} =
%\bar{H}_{\mathrm{S(T)}}(\bi{k}), \\
%& \bar{\Gamma}=\check{\Gamma}\, \hat{\tau}_0.
%\end{eqnarray}
%
The BdG Hamiltonian can be described in reduced $4 \times 4$ matrix form
\begin{eqnarray}
%\mathcal{H}_{0} =& \int d\bi{r} 
%\left[ \psi_{1,\uparrow}^\dagger(\bi{r}), \psi_{1,\downarrow}^\dagger(\bi{r}),
% \psi_{2,\uparrow}^\dagger(\bi{r}), \psi_{2,\downarrow}^\dagger(\bi{r}),
% \psi_{1,\uparrow}(\bi{r}), \psi_{1,\downarrow}(\bi{r}),
% \psi_{2,\uparrow}(\bi{r}), \psi_{2,\downarrow}(\bi{r})\right] \nonumber\\
%&\times \check{H}_{0}
%\left[
% \psi_{1,\uparrow}(\bi{r}), \psi_{1,\downarrow}(\bi{r}),
% \psi_{2,\uparrow}(\bi{r}), \psi_{2,\downarrow}(\bi{r}),
%\psi_{1,\uparrow}^\dagger(\bi{r}), \psi_{1,\downarrow}^\dagger(\bi{r}),
% \psi_{2,\uparrow}^\dagger(\bi{r}), \psi_{2,\downarrow}^\dagger(\bi{r}) 
% \right]^{\mathrm{T}},\\
% 
 \check{H}_{0}(\bi{k})=
 & \left[ \begin{array}{cccc} 
 \xi(\bi{k}) 
  & v\, e^{i\theta} & 0 & \Delta \\
v\, e^{-i\theta} & \xi(\bi{k}) & -s_s\Delta & 0 \\
0 & -s_s \Delta & -\xi(\bi{k})  & -v\, e^{-i\theta}\\
  \Delta  & 0 & -v\, e^{i\theta} & -\xi(\bi{k})
\end{array}\right], \label{hbdg}
\end{eqnarray}
by choosing spin of an electron as $\uparrow$ and that of a hole as $\downarrow$, 
where $s_s=1$ for a spin-triplet superconductor and $s_s=-1$ for a spin-singlet superconductor. 
We note in the normal state that $\xi^\ast(-\bi{k})=\xi(\bi{k})$ holds true in the presence of time-reversal symmetry.
%---------------------------------------------------------------------------

The Green function is obtained by solving the Gor'kov equation,
\begin{eqnarray}
&\left[ i\omega_n \check{1}- \check{H}_{0}(\bi{k}) \right] \check{G}_0(\bi{k},i\omega_n)=\check{1},\label{gorkov0}\\
& \check{G}_0(\bi{k},i\omega_n)
=\left[ \begin{array}{cc}
\hat{\mathcal{G}}_0(\bi{k},i\omega_n) & \hat{\mathcal{F}}_0(\bi{k},i\omega_n) \\
-s_s\hat{\mathcal{F}}_0^\ast(-\bi{k},i\omega_n) & -\hat{\mathcal{G}}_0^\ast(-\bi{k},i\omega_n) 
\end{array}\right],
\end{eqnarray}
where $\omega_n=(2n+1)\pi T$ is a fermionic Matsubara frequency with $T$ being a temperature.
The solution of the normal Green function within the first order of $\Delta$ 
is represented as
\begin{eqnarray}
\hat{\mathcal{G}}_0(\bi{k},\omega_n)&=& \frac{ \xi(\xi+2i\omega_n) - \omega_n^2 - v^2}{Z_0}
%\nonumber\\
%&\times& 
\left[ (i\omega_n-\xi) \hat{\rho}_0 + v \cos\theta \hat{\rho}_1 - v \sin\theta \hat{\rho}_2\right],\label{g0}\\
Z_0&=& \xi^4+2\xi^2(\omega_n^2-v^2) +(\omega_n^2+v^2)^2,
\end{eqnarray}
where we omit $\bi{k}$ from $\xi(\bi{k})$ for simplicity.
The results are common in both spin-singlet and spin-triplet cases because the normal Green function does not 
include the pair potential in the lowest order of $\Delta$.
The anomalous Green functions for a spin-singlet superconductor within the first order of $\Delta$ 
is calculated as
\begin{eqnarray}
\hat{\mathcal{F}}_0(\bi{k},\omega_n)&=& \frac{ \Delta}{Z_0}
 \left[ 2v \cos\theta \,\xi\,  \hat{\rho}_0 
%\right.\nonumber\\ &-&\left.
 -(\omega_n^2+v^2+\xi^2) \hat{\rho}_1 + 2i v \sin\theta \,\xi\,  \hat{\rho}_3\right],\label{f0_singlet}
\end{eqnarray}
The $\hat{\rho}_1$ component in Eq.~(\ref{f0_singlet}) is 
linked to the pair potential through the gap equation
\begin{eqnarray}
\Delta&=& - g\,  T\sum_{\omega_n} \frac{1}{V_{\mathrm{vol}}}\sum_{\bi{k}} \frac{1}{2} \mathrm{Tr}
[ \hat{\mathcal{F}}_0(\bi{k},\omega_n) \, \hat{\rho}_1 ], \label{gap_eq1}\\
&=&\pi\, g\, N_0\,   T \sum_{\omega_n} \frac{\Delta}{|\omega_n|},\label{gap_bcs}
\end{eqnarray}
where $N_0$ is the density of states at the Fermi level per spin. 
We have used the relation
\begin{eqnarray}
\frac{1}{V_{\mathrm{vol}}} \sum_{\bi{k}} \frac{a+b \,\xi^2}{Z_0}= 
\frac{\pi N_0 \left[a+b (\omega_n^2+v^2) \right]}{2\, |\omega_n|(\omega_n^2+v^2)},
\end{eqnarray}
where $a$ and $b$ are constants.
The last equation in Eq.~(\ref{gap_bcs}) is identical to the gap equation in the BCS theory.
The hybridization generates the $\hat{\rho}_0$ and $\hat{\rho}_3$ components in Eq.~(\ref{f0_singlet}) 
which belong to 
even-frequency spin-singlet even-momentum-parity even-band-parity (ESEE) symmetry class.

%The resulting gap equation in the dirty cace becomes Eq.~(\ref{gap_singlet}).
%Therefore, the hybridization does not affect $T_c$ in a spin-singlet superconductor.

In the case of a spin-triplet superconductor,
the anomalous Green function becomes
\begin{eqnarray}
\hat{\mathcal{F}}_0&(\bi{k},\omega_n)= \frac{ \Delta}{Z_0}
\left[   2  \omega_n\, v \sin\theta \, \hat{\rho}_0 
%\right.\nonumber \\
%&\left. 
-(\omega_n^2-v^2+\xi^2) i\, \hat{\rho}_2 - 2 i \omega_n\, v \,\cos\theta \hat{\rho}_3\right].\label{f0_triplet}
\end{eqnarray}
The $\hat{\rho}_2$ component is linked to the pair potential.
The gap equation is represented by Eq.~(\ref{gap_eq1}) with replacing $\hat{\rho}_1$ by $-i \hat{\rho}_2$.
%By integrating the anomalous Green function over $\bi{k}$, 
%
%\begin{eqnarray}
%\frac{1}{V_{\mathrm{vol}}}& \sum_{\bi{k}} \hat{\mathcal{F}}_0(\bi{k},\omega_n) = \frac{-\pi N_0}{|\omega_n|(\omega_n^2+v^2)}
%\nonumber\\
%\times&\left[ -\omega_n \, v\, \sin\theta\, \hat{\rho}_0 + \omega_n^2 \, i \, \hat{\rho}_2 + i\omega_n \, v\, \cos\theta \, \hat{\rho}_3 \right],
%\label{f0th_ave}
%\end{eqnarray}
%
The results of the gap equation in the linear regime, 
\begin{eqnarray}
\Delta =\pi\, g\, N_0\, T\, \sum_{\omega_n} \frac{\Delta\, |\omega_n|}{\omega_n^2+v^2}. \label{gap_triplet_clean}
\end{eqnarray}
 deviate from Eq.~(\ref{gap_bcs}). 
In Eq.~(\ref{f0_triplet}), the hybridization generates the $\hat{\rho}_0$ and $\hat{\rho}_3$ components which belong to 
odd-frequency spin-triplet even-momentum-parity even-band-parity (OTEE) symmetry 
class~\cite{BSchaffer:prb2013,asano15,vasenko:jpcm2017,vasenko:jetplett2017}.
The hybridization suppresses $T_c$ because an odd-frequency pair is
 thermodynamically unstable~\cite{asano15,asano11,suzuki:prb2015}.
At $v=0$, the gap equation in Eq.~(\ref{gap_triplet_clean}) is identical 
to Eq.~(\ref{gap_bcs}) because the odd-frequency pairing correlations are absent.

%=========================================================================
\section{Effects of impurities}\label{impurity}
%=========================================================================

Let us consider the nonmagnetic random impurities described by 
\begin{eqnarray}
\check{H}_{\mathrm{imp}} &=&V_{\mathrm{imp}}(\bi{r})\left[ \begin{array}{cccc}
1 & e^{i\theta} & 0 & 0 \\
e^{-i\theta} & 1& 0 & 0 \\
0 & 0 & -1 & - e^{-i\theta} \\
0 & 0 & - e^{i\theta} & -1
\end{array}\right],\\
&=& V_{\mathrm{imp}}(\bi{r}) \hat{\tau}_3\, \hat{\rho}_0 +
V_{\mathrm{imp}}(\bi{r}) \check{A}, \label{vimp0}\\
 \check{A}&=&  \hat{\tau}_3\, \hat{\rho}_1
\, \cos\theta - \hat{\rho}_2 \, \sin{\theta}.
\end{eqnarray}
The first and the second terms in Eq.~(\ref{vimp0}) cause the intraband and the interband scatterings, respectively.
We assume that the impurity potential satisfies the following properties,
\begin{eqnarray}
\overline{V_{\mathrm{imp}}(\bi{r})}&=&0,\label{vimp_1}\\
\overline{ V_{\mathrm{imp}}(\bi{r}) V_{\mathrm{imp}}(\bi{r}') }& =& n_{\mathrm{imp}} v_{\mathrm{imp}}^2
\delta(\bi{r}-\bi{r}'), \label{vimp_2}
\end{eqnarray}
where $\overline{\cdots}$ means the ensemble average, $n_{\mathrm{imp}}$ is the impurity concentration, and $v_{\mathrm{imp}}$ 
represents the strength of the impurity potential. 
We also assume that the attractive electron-electron interactions are insensitive to 
the impurity potentials~\cite{anderson:jpcs1959}.
To discuss the effects of impurities with Eqs.~(\ref{vimp_1}) and (\ref{vimp_2}), 
Hamiltonian in real space is necessary.
The impurity Hamiltonian in Eq.~(\ref{vimp0}) is described in real space as well 
as the kinetic part and the hybridization in 
Eqs.~(\ref{hn_operator}) and (\ref{hn}). 
In the real space representation with the basis shown in Eq.~(\ref{hn_operator}), the random potential 
$V_{\mathrm{imp}}(\boldsymbol{r})$ should be independent of band indices.
The phase of random potential generating the interband scattering must be equal to that of the hybridization. 
Otherwise, time-reversal symmetry is broken.
The effects of the impurity scatterings are taken into account through the self-energy 
estimated within the Born approximation.
The Green function in the presence of the impurity potential is calculated 
within the second order perturbation expansion with respect to the impurity potential,%
\begin{eqnarray}
 &\check{{G}}(\bi{r}-\bi{r}', \omega_n) \approx 
\check{{G}}_0(\bi{r}-\bi{r}', \omega_n) 
%\nonumber\\
%
%&
+\int d\bi{r}_1 \check{{G}}_0(\bi{r}-\bi{r}_1, \omega_n)\, 
\overline{\check{H}_{\mathrm{imp}}(\bi{r}_1)}\, 
\check{{G}}(\bi{r}_1-\bi{r}', \omega_n) \nonumber\\
&+\int d\bi{r}_1 \int d\bi{r}_2 \, 
\check{{G}}_0(\bi{r}-\bi{r}_1, \omega_n)\, 
%\nonumber\\
%&\times 
\overline{\check{H}_{\mathrm{imp}}(\bi{r}_1)\, 
\check{{G}}_0(\bi{r}_1-\bi{r}_2, \omega_n)\, \check{H}_{\mathrm{imp}}(\bi{r}_2)}\nonumber\\ 
&\times \check{{G}}(\bi{r}_2-\bi{r}', \omega_n),
\end{eqnarray}
where 0 in the subscript indicates unperturbed Green function.
By considering Eqs.~(\ref{vimp_1}) and (\ref{vimp_2}), we obtain
\begin{eqnarray}
 &\check{{G}}&(\bi{r}-\bi{r}', \omega_n) = 
\check{{G}}_0(\bi{r}-\bi{r}', \omega_n) \nonumber\\
&+&n_{\mathrm{imp}}\, v_{\mathrm{imp}}^2 \int d\bi{r}_1 
\check{{G}}_0(\bi{r}-\bi{r}_1, \omega_n)\,  \hat{\tau}_3 \,
\check{{G}}_0(0, \omega_n) \, \hat{\tau}_3 \, 
\check{{G}}(\bi{r}_1-\bi{r}', \omega_n) \nonumber\\
&+&n_{\mathrm{imp}}\, v_{\mathrm{imp}}^2 \int d\bi{r}_1 
\check{{G}}_0(\bi{r}-\bi{r}_1, \omega_n)  \, \check{A} \,
\check{{G}}_0(0, \omega_n) \,  \check{A} \, 
\check{{G}}(\bi{r}_1-\bi{r}', \omega_n). \label{expansion2}
\end{eqnarray}
The second and the third terms are derived from the intraband impurity scatterings and the interband 
impurity scatterings, respectively. 
% Here we introduce $v_a$ and $v_b$ to distinguish 
%the two self-energies. We will put $v_a=v_b=v_{\mathrm{imp}}$ afterword. 
%The term proportional to $\hat{\rho}_1 \hat{\tau}_3 \check{{G}}_0 \hat{\tau}_3$ 
%and that proportional to $ \hat{\tau}_3 \check{{G}}_0 \hat{\rho}_1 \hat{\tau}_3$
%do not appear because the final state after applying the second order 
%perturbation expansion should be identical to the initial state 
%in the Born approximation. 
By applying the Fourier transformation, the Green function becomes
\begin{eqnarray}
 \check{{G}}(\bi{k},\omega_n) &= &
\check{{G}}_0(\bi{k},\omega_n)
% \nonumber\\
%&+& 
+\check{{G}}_0(\bi{k},\omega_n) \, 
\check{\Sigma}_{\mathrm{imp}}(\omega_n) \, \check{{G}}(\bi{k},\omega_n),\label{expansion3}\\
\check{\Sigma}_{\mathrm{imp}}&=&\check{\Sigma}_{\mathrm{intra}} + \check{\Sigma}_{\mathrm{inter}}, \label{sigma_imp_def}
\end{eqnarray}
where $\check{\Sigma}_{\mathrm{intra}}$ and $\check{\Sigma}_{\mathrm{inter}}$ are the self-energy due to the 
intraband impurity scatterings and that of interband impurity scatterings, respectively. 
The details of the derivation are given in Appendix.
In the Born approximation, 
the self-energies are represented as
\begin{eqnarray}
\check{\Sigma}_{\mathrm{intra}} &=& n_{\mathrm{imp}} v_{\mathrm{imp}}^2 \hat{\tau}_3\, \hat{\rho}_0 
\frac{1}{V_{\mathrm{vol}}} \sum_{\bi{k}} \check{G}_0(\bi{k},\omega_n)
\hat{\tau}_3\, \hat{\rho}_0,\label{sigma_intra}\\
\check{\Sigma}_{\mathrm{inter}} &=& n_{\mathrm{imp}} v_{\mathrm{imp}}^2 \, \check{A}  \,
\frac{1}{V_{\mathrm{vol}}} \sum_{\bi{k}} \check{G}_0(\bi{k},\omega_n) \, \check{A}. \label{sigma_inter}
\end{eqnarray}
The total self-energy is calculated as
\begin{eqnarray}
\check{\Sigma}_{\mathrm{imp}} 
&=&\left[ \begin{array}{cc}
\hat{\Sigma}_G & \hat{\Sigma}_F \\
-s_s \hat{\Sigma}_F^\ast & - \hat{\Sigma}_G^\ast
\end{array}\right],\label{sigma_imp}
\end{eqnarray}
with
\begin{eqnarray}
\hat{\Sigma}_G &= &2 n_{\mathrm{imp}} v_{\mathrm{imp}}^2 
%\nonumber\\
%&\times&
\left[
\langle g_0 \rangle \hat{\rho}_0 + \cos\theta\, S_g\, \hat{\rho}_1
- \sin\theta \, S_g\, \hat{\rho}_2 \right],
\label{sigg}\\
\hat{\Sigma}_F &=& -2 n_{\mathrm{imp}} v_{\mathrm{imp}}^2
%\nonumber\\
%&\times&
\left[
\cos\theta\, S_f\, \hat{\rho}_0 
+ \langle f_1 \rangle  \hat{\rho}_1 
%\right.\nonumber\\
%&\left. 
+i \sin\theta\, S_f\, \hat{\rho}_3 \right],
\label{sigf}\\
S_g &=& \langle g_1 \rangle \cos\theta - \langle g_2 \rangle \sin\theta,\\
S_f &=& \langle f_0 \rangle \cos\theta - i \langle f_3 \rangle \sin\theta.
\end{eqnarray}
Here the Green function after carrying out the summation of $\bi{k}$ 
is indicated by $\langle \cdots \rangle$  as,
\begin{eqnarray}
\langle \hat{\mathcal{G}}_0(\omega_n) \rangle\equiv &
\frac{1}{V_{\mathrm{vol}}} \sum_{\bi{k}} \hat{\mathcal{G}}_0(\bi{k},\omega_n) 
=\sum_{\nu=0}^3 \langle g_\nu \rangle \hat{\rho}_\nu, \label{ksum_g}\\
\langle \hat{\mathcal{F}}_0(\omega_n) \rangle\equiv & \frac{1}{V_{\mathrm{vol}}} \sum_{\bi{k}} 
\hat{\mathcal{F}}_0(\bi{k},\omega_n) 
 =\sum_{\nu=0}^3 \langle f_\nu \rangle \hat{\rho}_\nu, \label{ksum_f}
\end{eqnarray}
where $\hat{\rho}_\nu$ with $\nu=0-3$ are the Pauli matrices in band space.
The Gor'kov equation in the presence of impurities is expressed by
\begin{eqnarray}
&\left[ i\omega_n \check{1}- \check{H}_{0}(\bi{k}) -\check{\Sigma}_{\mathrm{imp}}\right] 
\check{G}(\bi{k},i\omega_n)=\bar{1},\label{gorkov1}\\
& \check{G}(\bi{k},i\omega_n)
=\left[ \begin{array}{cc}
\hat{\mathcal{G}}(\bi{k},i\omega_n) & \hat{\mathcal{F}}(\bi{k},i\omega_n) \\
-s_s\hat{\mathcal{F}}^\ast(-\bi{k},i\omega_n) & -\hat{\mathcal{G}}^\ast(-\bi{k},i\omega_n) 
\end{array}\right],
\end{eqnarray}
Eq.~(\ref{sigma_imp}) with Eqs.~(\ref{sigg})-(\ref{ksum_f}) give the 
general expression self-energy due to impurity scattering within the Born approximation. 
The properties in the normal state and those in the superconducting state are mainly 
embedded in the normal Green function in Eq.~(\ref{ksum_g}) and in the anomalous Green function 
in Eq.~(\ref{ksum_f}), respectively.
Therefore the results can be applied to various two-band superconductors.
Here we briefly mention a general feature of the self-energy. 
In Eq.~(\ref{sigf}), $\langle f_1\rangle \hat{\rho}_1$ is present but
$\langle f_2 \rangle \hat{\rho}_2$ is absent in $\hat{\Sigma}_F$ because of the anticommutation relations
among $\hat{\rho}_\nu$. This feature is independent of the normal state Hamiltonian.
As shown in the remaining part of this section, the effects of random nonmagnetic impurity scatterings 
on the transition temperature $T_c$ depends on spin symmetry of the pair potential. 
The difference comes from such general property of $\hat{\Sigma}_F$. 
We will explain details of the difference in the following subsections.

%-----------------------------------------------------
\subsection{spin-singlet}
%-----------------------------------------------------
The normal part of the self-energy is calculated as
\begin{eqnarray}
\hat{\Sigma}_G &=& \frac{-i\omega_n}{2\tau_{\mathrm{imp}}|\omega_n|} \hat{\rho}_0, \label{sigg2}\\
\frac{1}{\tau_{\mathrm{imp}}}&=& 2\times 2\pi N_0 n_{\mathrm{imp}} v_{\mathrm{imp}}^2, \label{tauimp_def}
\end{eqnarray}
where $\tau_{\mathrm{imp}}$ represents the life time due to impurity scatterings.
The factor 2 in Eq.~(\ref{tauimp_def}) stems from the two contributions of 
different scattering processes: the 
intraband impurity scatterings and the interband impurity scatterings. 
In a spin-singlet superconductor,
the self-energy of the anomalous part results in
\begin{eqnarray}
\hat{\Sigma}_F = \frac{\Delta}{2\tau_{\mathrm{imp}}|\omega_n|} \hat{\rho}_1, \label{sigf_singlet}
\end{eqnarray}
because Eq.~(\ref{sigf}) includes $\langle f_1\rangle \hat{\rho}_1$.
% 
%and
%\begin{eqnarray}
%\langle \check{G}_0(\omega_n) \rangle = 
%-\frac{N_0 \pi}{|\omega_n|}\left[\begin{array}{cccc}
%i\omega_n & 0 &0 & \Delta \\
%0 & i\omega_n & \Delta & 0\\
%0 & \Delta & i\omega_n & 0 \\
%\Delta &0 & 0 & i\omega_n
%\end{array}\right].\label{ksum_g44_singlet}
%\end{eqnarray}
%The summation over momenta can be done as shown in Appendix~\ref{ksum}. 
%With Eq.~(\ref{f0_ave_singlet}), the anomalous part of the self-energy is calculated as
%
%\begin{eqnarray}
%\hat{\Sigma}_F=&\frac{\Delta\,  \hat{\rho}_1}{4\tau_{\mathrm{imp}}|\omega_n|} + \hat{\rho}_1\, 
%\frac{\Delta\,  \hat{\rho}_1}{4\tau_{\mathrm{imp}}|\omega_n|} \, \hat{\rho}_1
%= \frac{\Delta}{2\tau_{\mathrm{imp}}|\omega_n|} \hat{\rho}_1,\label{sigmaf_singlet}\\
%\frac{1}{\tau_{\mathrm{imp}}}=& 2\times 2\pi N_0 n_{\mathrm{imp}} v_{\mathrm{imp}}^2, \label{tauimp_def}
%\frac{1}{\tau_{\mathrm{imp}}} = \frac{1}{\tau_a} + \frac{1}{\tau_b}.
%\end{eqnarray}
%Together with the Fermi velocity $v_F= k_F/m$, we define the mean free path as $\ell=v_F \tau_{\mathrm{imp}}$.
As a consequence, the Gor'kov equation in the presence of impurities becomes,
\begin{eqnarray}
&\left[\begin{array}{cc}
(i\tilde{\omega}_n-\xi)\hat{\rho}_0 - \hat{V}& - \tilde{\Delta} \hat{\rho}_1 \\
- \tilde{\Delta} \hat{\rho}_1 & (i\tilde{\omega}_n+\xi)\hat{\rho}_0  + \hat{V}^\ast
\end{array}\right]
\, \check{G}(\bi{k},i\omega_n) = \check{1},\label{gorkov_singlet}\\
& \hat{V}= v\cos\theta \, \hat{\rho}_1 - v \sin\theta \, \hat{\rho}_2,\\
%& \check{G}(\bi{k},i\omega_n)
%=\left[ \begin{array}{cc}
%\hat{\mathcal{G}}(\bi{k},i\omega_n) & \hat{\mathcal{F}}(\bi{k},i\omega_n) \\
%\hat{\mathcal{F}}^\ast(-\bi{k},i\omega_n) & -\hat{\mathcal{G}}^\ast(-\bi{k},i\omega_n) 
%\end{array}\right],\\
& \tilde{\omega}_n=\omega_n\, \eta_n, \quad \tilde{\Delta}=\Delta\, \eta_n, \quad 
\eta_n= 1+\frac{1}{2\tau_{\mathrm{imp}}|\omega_n|}.
\end{eqnarray}
The self-energy renormalizes 
the frequency and the pair potential exactly in the same manner as $\omega_n \to \tilde{\omega}_n$ 
and $\Delta \to \tilde{\Delta}$. 
As a consequence, the anomalous Green function can be calculated as
\begin{eqnarray}
\hat{\mathcal{F}}(\bi{k}, \omega_n) = \left.\hat{\mathcal{F}}_0
(\bi{k}, \tilde{\omega}_n)\right|_{\Delta \to \tilde{\Delta}}, \label{ftilde_singlet}
\end{eqnarray}
where $\hat{\mathcal{F}}_0$ on the right hand side is shown in Eq.~(\ref{f0_singlet}).
The gap equation in the presence of impurities is given by Eq.~(\ref{gap_eq1}) 
with $\hat{\mathcal{F}}_0(\bi{k}, \omega_n) \to \hat{\mathcal{F}}(\bi{k}, \omega_n)$.  
The resulting gap equation 
\begin{eqnarray}
\Delta=& 
=&\pi\, g\,  N_0\,   T \sum_{\omega_n} \frac{\tilde{\Delta}}{|\tilde{\omega}_n|}=\pi \, g\, N_0\,  T \sum_{\omega_n} 
\frac{{\Delta}}{|{\omega}_n|}, \label{gap_singlet}
\end{eqnarray}
remains unchanged from that in the clean limit. Thus the impurity scatterings do not change $T_c$ 
in a spin-singlet superconductor.
The argument here is exactly the same as that in Ref.~\cite{agd} for a single-band spin-singlet 
$s$-wave superconductor and is
consistent with the Anderson's theorem~\cite{anderson:jpcs1959}.
%The Green function after the summation over $\bi{k}$ describes physics in a dirty superconductor.
%Actually, we find in Eq.~(\ref{ksum_g44_singlet}) that 
%$\langle \check{G}_0(\omega_n) \rangle$ consists of two copies of the Green function 
%for a single-band spin-singlet $s$-wave superconductor.

%-------------------------------------------------
\subsection{spin-triplet}
%-------------------------------------------------
In a spin-triplet superconductor, 
the Green function in Eq.~(\ref{ksum_f}) with Eq.~(\ref{f0_triplet}) is calculated as
\begin{eqnarray}
%\langle \hat{\mathcal{G}}_0 \rangle = \frac{N_0 \pi}{|\omega_n|} (-i \omega_n) \hat{\rho}_0, \quad
\langle \hat{\mathcal{F}}_0 \rangle = \frac{N_0 \pi \Delta}{|\omega_n|(\omega_n^2+v^2)} 
\left[ \omega_n\, v\, \sin\theta \,\hat{\rho}_0 - i \,\omega_n^2 \hat{\rho}_2
-i\, \omega_n \, v\, \cos\theta \, \hat{\rho}_3 \right].
\label{f0_ave_triplet}
\end{eqnarray}
%and
% 
%\begin{eqnarray}
%\langle \check{G}_0(\omega_n) \rangle = 
%-\frac{N_0 \pi}{|\omega_n|}\left[\begin{array}{cccc}
%i\omega_n & 0 &0 & \Delta \\
%0 & i\omega_n & -\Delta & 0\\
%0 & -\Delta & i\omega_n & 0 \\
%\Delta &0 & 0 & i\omega_n
%\end{array}\right].\label{ksum_g44_triplet}
%\end{eqnarray}
%The anomalous Green function after summation of $\bi{k}$ 
%has the $\hat{\rho}_2$ component. 
By substituting the results into Eq.~(\ref{sigf}), we find
\begin{eqnarray}
\hat{\Sigma}_F=0, \label{sigf_triplet}
\end{eqnarray}
because Eq.~(\ref{sigf}) does not include $\langle f_2 \rangle \hat{\rho}_2$.
The resulting Gor'kov equation becomes,
\begin{eqnarray}
&\left[\begin{array}{cc}
(i\tilde{\omega}_n-\xi)\hat{\rho}_0 -\hat{V} & -{\Delta} i \hat{\rho}_2 \\
 {\Delta} i \hat{\rho}_2 & (i\tilde{\omega}_n+\xi)\hat{\rho}_0 +\hat{V}^\ast
\end{array}\right]
 \check{G}(\bi{k},i\omega_n) = \check{1},\label{gorkov_triplet}
%& \check{G}(\bi{k},i\omega_n)
%=\left[ \begin{array}{cc}
%\hat{\mathcal{G}}(\bi{k},i\omega_n) & \hat{\mathcal{F}}(\bi{k},i\omega_n) \\
%-\hat{\mathcal{F}}^\ast(-\bi{k},i\omega_n) & -\hat{\mathcal{G}}^\ast(-\bi{k},i\omega_n) 
%\end{array}\right].
\end{eqnarray}
The impurity self-energy renormalizes the frequency as $\omega_n\to \tilde{\omega}_n$ but leaves the pair potential 
as it is.
Thus the anomalous Green function in the presence of impurities becomes
\begin{eqnarray}
\hat{\mathcal{F}}(\bi{k}, \omega_n) = \hat{\mathcal{F}}_0
(\bi{k}, \tilde{\omega}_n), \label{ftilde_triplet}
\end{eqnarray}
where $\hat{\mathcal{F}}_0$ on the right hand side is given in Eq.~(\ref{f0_triplet}).
The gap equation Eq.~(\ref{gap_eq1}) 
with $\hat{\mathcal{F}}_0(\bi{k}, \omega_n) \to \hat{\mathcal{F}}(\bi{k}, \omega_n)$ and 
$\hat{\rho}_1\to -i \hat{\rho}_2$ results in
\begin{eqnarray}
\Delta
%=&\pi g N_0  T \sum_{\omega_n} \frac{ {\Delta} }{|\tilde{\omega}_n|}
= \pi g N_0 T \sum_{\omega_n} 
\frac{\Delta \, (|{\omega}_n| +1/ 2\tau_{\mathrm{imp}})}{ (|{\omega}_n| +1/ 2\tau_{\mathrm{imp}})^2 + v^2 }. \label{gap_triplet}
\end{eqnarray}
The results suggest that the impurity scatterings decrease $T_c$ for a spin-triplet superconductor.
%The gap equation in Eq.~(\ref{gap_triplet}) is exactly the same as that of single-band unconventional 
%superconductors characterize=0$ such symmetry as $p$- and $d$-wave. 

\begin{figure}[tbh]
\begin{center}
\includegraphics[width=10.5cm]{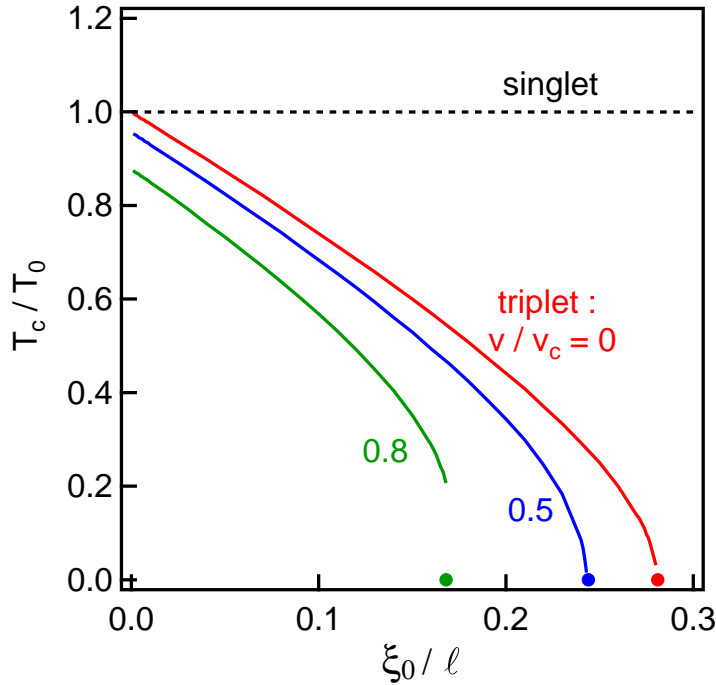}
\end{center}
\caption{ 
The transition temperature $T_c$ versus $\xi_0/\ell$. The impurity concentration is 
proportional to $\xi_0/\ell$, where $\xi_0$ is the coherence length and $\ell$ is the elastic mean free path.
In a spin-singlet case, $T_c$ is independent of $\xi_0/\ell$ within the Born approximation as shown with 
a broken line, which is consistent with the Anderson's theorem.
The results for a spin-triplet interband superconductor at $v=0$ are identical to those for a single-band 
unconventional superconductor characterized such symmetry as spin-singlet $d$-wave or spin-triplet $p$-wave.
}
\label{fig:tc}
\end{figure}

In Fig.~\ref{fig:tc}, we show $T_c$ of a spin-triplet interband superconductor as a function of $\xi_0/\ell$, where 
$T_0$ is the transition temperature in the clean limit in the absence of the hybridization (i.e., $v=0$), 
$\xi_0=v_F/2\pi T_0$ is the coherence length, $v_F=k_F/m$ is the Fermi velocity, and
$\ell=v_F \, \tau_{\mathrm{imp}}$ is the mean free path due to the impurity scatterings. 
We numerically solve Eq.~(\ref{gap_triplet}) with $\omega_c/2\pi T_0 = 10^3$.
The results show that $T_c$ decreases with the increase of $\xi_0/\ell$.
In the clean limit, $T_c$ decreases with the increase of the hybridization $v$ as indicated in
Eq.~(\ref{gap_triplet_clean}).
The superconducting phase vanishes when the amplitude of hybridization goes over 
its critical value of $v_c \approx 2\pi T_0 / C$, where 
$ C=4\, e^{\gamma_E} $ and  $\gamma_E=0.577$ is the Euler's constant. 
In the presence of impurities, the interband spin-triplet superconductivity 
vanishes at $\xi_0/\ell \approx 2/C = 0.281$ at $v=0$, 
 $\xi_0/\ell \approx 0.244$ at $v=0.5\, v_c$, and  $\xi_0/\ell \approx 0.168$ at $v=0.8\, v_s$.

The suppression of $T_c$ by impurities in a spin-triplet case can be interpreted as follows. 
The interband impurity scatterings hybridize the electronic states in the two bands and average 
the pair potential over the two-band degree of freedom.
As shown in Eq.~(\ref{hbdg}), the sign of pair potential in one sector is opposite to that in the other 
where we set $s_s=1$ for a triplet superconductor. 
Thus the pair potentials in the two sectors cancel each other 
when the interband impurity potential hybridizes the two sectors.
As a result, the anomalous part of the self-energy vanishes as shown in Eq.~(\ref{sigf_triplet}).
Namely, the impurity self-energy does not renormalize the pair potential, 
which leads to the suppression of $T_c$. 
 The absence of 
$\langle f_2\rangle \hat{\rho}_2$ in Eq.~(\ref{sigf}) can be understood by such physical interpretation. 
It would be worth mentioning that the gap equation in Eq.~(\ref{gap_triplet}) with $v=0$ is 
identical to that for a single-band unconventional superconductor under the potential disorder.
In a $p$-wave or $d$-wave superconductor, the anomalous Green function $\langle \mathcal{F}_0(\omega_n) \rangle$ 
vanishes due to their unconventional pairing symmetries, which leads to $\Sigma_F=0$ and the suppression of $T_c$. 
We conclude that the odd-band-parity pairing correlation is fragile 
under impurity potential even though it belongs to $s$-wave momentum parity symmetry class.
Therefore, a clean enough sample is necessary to observe spin-triplet 
interband superconductivity in experiments. 

Mathematically, the robustness of a spin-singlet $s$-wave interband superconducting state is  
described by the anomalous part of the self-energy $\hat{\Sigma}_F= \hat{\Delta}/2\tau_{\mathrm{imp}}|\omega_n|$ 
in Eq.~(\ref{sigf_singlet}). 
The suppression of $T_c$ in a spin-triplet superconductor is described by 
$\hat{\Sigma}_F=0$ in Eq.~(\ref{sigf_triplet}). 
As we already explained below Eq.~(\ref{gorkov1}), these features are derived from 
the general expression of the self-energy in Eq.~(\ref{sigf}) and are 
independent of the normal state Hamiltonian.
Therefore, our conclusions are valid for various interband superconductors.

%***************************************************************
\section{Conclusion}
%***************************************************************
We studied the effects of random nonmagnetic impurities on the superconducting transition temperature $T_c$
in a two-band superconductor characterized by an equal-time $s$-wave interband pair potential.
Due to the two-band degree of freedom, both spin-singlet and spin-triplet pairing order parameters 
satisfy the requirement from the Fermi-Dirac statistics of electrons.
The effects of impurity potential is considered through the self-energy obtained within the Born approximation.
The transition temperature is calculated from the linearized gap equation.
In a spin-singlet superconductor, the random potential does not change $T_c$.
On the other hand in a spin-triplet superconductor, $T_c$ decreases with the increase of the impurity concentration.
We conclude that Cooper pairs belonging to odd-band-parity symmetry class 
are fragile under the random impurity potential even though they belong to $s$-wave momentum symmetry.

\section*{Acknowledgments}
The authors are grateful to Y. Tanaka, and Ya. V. Fominov for useful discussions.
This work was supported by Topological Materials Science (Nos.~JP15H05852 and JP15K21717) and 
KAKENHI (No.~JP15H03525) from the Ministry of Education, Culture, Sports, Science and Technology (MEXT) of 
Japan, JSPS Core-to-Core Program (A. Advanced Research Networks), Japanese-Russian JSPS-RFBR project 
(Nos.~2717G8334b and 17-52-50080), and by the Ministry of Education and Science of the Russian Federation
(Grant No.~14Y.26.31.0007).

\section*{Appendix}
We show the details of the derivation of the impurity self-energy in Eq.~(\ref{sigma_imp}).
The Fourier representation of the Green function is defined by
\begin{equation}
\check{G}(\bi{r}-\bi{r}', \omega_n) = \frac{1}{V_{\mathrm{vol}}} \sum_{\bi{k}}
\check{G}(\bi{k}, \omega_n) e^{i \bi{k} \cdot (\bi{r}-\bi{r}')}.\label{fourier_g}
\end{equation}
The Green function $ \check{G}_0(0, \omega_n)$ in Eq.~(\ref{expansion2}) 
is obtained by putting $\bi{r}=\bi{r}'$. 
When we substitute Eq.~(\ref{fourier_g}) into Eq.~(\ref{expansion2}) and carrying out the integration 
over $\bi{r}_1$, we find Eq.~(\ref{expansion3}). Since $\check{G}_0(\bi{k},\omega_n)$ satisfies Eq.~(\ref{gorkov0}), 
we obtain Eq.~(\ref{gorkov1}) with the self-energy in Eq.~(\ref{sigma_imp_def}).
To proceed the calculation, the Green function integrated over the momenta is necessary. 
The general expression of them are defined by Eqs.~(\ref{ksum_g}) and (\ref{ksum_f}). 
By substituting Eqs.~(\ref{ksum_g}) and (\ref{ksum_f}) into Eqs.~(\ref{sigma_intra}) and (\ref{sigma_inter}),
we find 
\begin{eqnarray}
\check{\Sigma}_{\mathrm{intra}}=&n_{\mathrm{imp}}\, v_{\mathrm{imp}}^2 \sum_{\nu=0}^3 
\left[\begin{array}{cc}
\langle g_\nu \rangle \hat{\rho}_\nu & -\langle f_\nu \rangle \hat{\rho}_\nu \\
s_s[\langle f_\nu \rangle \hat{\rho}_\nu]^\ast & -[\langle g_\nu \rangle \hat{\rho}_\nu]^\ast
\end{array}\right],\\
\check{\Sigma}_{\mathrm{inter}}=&n_{\mathrm{imp}}\, v_{\mathrm{imp}}^2 \sum_{\nu=0}^3 
\left[\begin{array}{cc}
\hat{A}_-\, \langle g_\nu \rangle \hat{\rho}_\nu \, \hat{A}_- & -\hat{A}_- \, \langle f_\nu \rangle \hat{\rho}_\nu \,\hat{A}_+\\
s_s\hat{A}_+[\langle f_\nu \rangle \hat{\rho}_\nu]^\ast \hat{A}_- & -\hat{A}_+[\langle g_\nu \rangle \hat{\rho}_\nu]^\ast\hat{A}_+
\end{array}\right],\\
\hat{A}_\pm =& \hat{\rho}_1\cos\theta\pm \hat{\rho}_2\sin\theta.
\end{eqnarray}
Here we focus on the anomalous part of the self-energy because 
its general expression is important to justify the main conclusion.
We find the relation,
\begin{eqnarray}
\hat{A}_-\, \sum_\nu \langle f_\nu \rangle \hat{\rho}_\nu \, \hat{A}_+
=& \langle f_1 \rangle \hat{\rho}_1 - \langle f_2 \rangle \hat{\rho}_2
+(\cos 2\theta \langle f_0 \rangle -i \sin 2\theta \langle f_3 \rangle) \hat{\rho}_0 \nonumber\\
&-(\cos 2\theta \langle f_3 \rangle -i \sin 2\theta \langle f_0 \rangle) \hat{\rho}_3.
\end{eqnarray}
The most important feature is that $\langle f_2 \rangle \hat{\rho}_2$ component changes its sign
due to the anticomutation relations among $\hat{\rho}_j$.
Together with the intraband contribution $\sum_\nu \langle f_\nu \rangle \hat{\rho}_\nu$, we
obtain the general expression of the anomalous part in Eq.~(\ref{sigf}).

\section*{References}

\providecommand{\newblock}{}


\begin{thebibliography}{10}
\expandafter\ifx\csname url\endcsname\relax
  \def\url#1{{\tt #1}}\fi
\expandafter\ifx\csname urlprefix\endcsname\relax\def\urlprefix{URL }\fi
\providecommand{\eprint}[2][]{\url{#2}}

\bibitem{agd}
Abrikosov A~A, Gor{\textquoteright}kov L~P and Dzyaloshinski I~E 1975 {\em
  {Methods of Quantum Field Theory in Statistical Physics}\/} (New York: Dover
  Publications)

\bibitem{abrikosov:jetp1959}
Abrikosov A~A and Gor{\textquoteright}kov L~P 1959 {\em Sov. Phys. JETP\/} {\bf
  9} 220

\bibitem{anderson:jpcs1959}
Anderson P 1959 {\em Journal of Physics and Chemistry of Solids\/} {\bf 11} 26
  -- 30 ISSN 0022-3697
  \urlprefix\url{http://www.sciencedirect.com/science/article/pii/0022369759900368}

\bibitem{allen:review1982}
Allen P~B and Mitrovic B 1983 Theory of superconducting tc ({\em Solid State
  Physics\/} vol~37) ed Ehrenreich H, Seitz F and Turnbull D (Academic Press)
  pp 1 -- 92
  \urlprefix\url{http://www.sciencedirect.com/science/article/pii/S0081194708606657}

\bibitem{golubov:prb1997}
Golubov A~A and Mazin I~I 1997 {\em Phys. Rev. B\/} {\bf 55}(22) 15146--15152
  \urlprefix\url{https://link.aps.org/doi/10.1103/PhysRevB.55.15146}

\bibitem{onari:prl2009}
Onari S and Kontani H 2009 {\em Phys. Rev. Lett.\/} {\bf 103} 177001--4

\bibitem{efremov:prb2011}
Efremov D~V, Korshunov M~M, Dolgov O~V, Golubov A~A and Hirschfeld P~J 2011
  {\em Phys. Rev. B\/} {\bf 84} 180512--4

\bibitem{korshunov:prb2014}
Korshunov M~M, Efremov D~V, Golubov A~A and Dolgov O~V 2014 {\em Phys. Rev.
  B\/} {\bf 90}(13) 134517
  \urlprefix\url{https://link.aps.org/doi/10.1103/PhysRevB.90.134517}

\bibitem{hoyer:prb2015}
Hoyer M, Scheurer M~S, Syzranov S~V and Schmalian J 2015 {\em Phys. Rev. B\/}
  {\bf 91}(5) 054501
  \urlprefix\url{https://link.aps.org/doi/10.1103/PhysRevB.91.054501}

\bibitem{asano172bandimp}
Asano Y and Golubov A~A 2017 Green function theory of dirty two-band
  superconductivity (\textit{Preprint} \eprint{arXiv:1710.04348})

\bibitem{mgb2:akimitsu2001}
Nagamatsu J, Nakagawa N, Muranaka T, Zenitani Y and Akimitsu J 2001 {\em Nature
  (London)\/} {\bf 410} 63

\bibitem{mgb2:louie2002}
Choi H~J, Roundy D, Sun H, Cohen M~L and Louie S~G 2002 {\em Nature (London)\/}
  {\bf 418} 758

\bibitem{pnictide:hosono2008}
Kamihara Y, Watanabe T, Hirano M and Hosono H 2008 {\em J. Am. Chem. Soc.\/}
  {\bf 130} 3296

\bibitem{hosono:physicac2015}
Hosono H and Kuroki K 2015 {\em Physica C: Superconductivity and its
  Applications\/} {\bf 514} 399 -- 422 ISSN 0921-4534 superconducting
  Materials: Conventional, Unconventional and Undetermined
  \urlprefix\url{http://www.sciencedirect.com/science/article/pii/S0921453415000477}

\bibitem{hor:prl2010}
Hor Y~S, Williams A~J, Checkelsky J~G, Roushan P, Seo J, Xu Q, Zandbergen H~W,
  Yazdani A, Ong N~P and Cava R~J 2010 {\em Phys. Rev. Lett.\/} {\bf 104}(5)
  057001
  \urlprefix\url{https://link.aps.org/doi/10.1103/PhysRevLett.104.057001}

\bibitem{fu:prl2010}
Fu L and Berg E 2010 {\em Phys. Rev. Lett.\/} {\bf 105}(9) 097001
  \urlprefix\url{https://link.aps.org/doi/10.1103/PhysRevLett.105.097001}

\bibitem{sato:repprogphys2017}
Sato M and Ando Y 2017 {\em Reports on Progress in Physics\/} {\bf 80} 076501
  \urlprefix\url{http://stacks.iop.org/0034-4885/80/i=7/a=076501}

\bibitem{upt3:rmp2002}
Joynt R and Taillefer L 2002 {\em Rev. Mod. Phys.\/} {\bf 74}(1) 235--294
  \urlprefix\url{https://link.aps.org/doi/10.1103/RevModPhys.74.235}

\bibitem{yanase:prb2016}
Yanase Y 2016 {\em Phys. Rev. B\/} {\bf 94}(17) 174502
  \urlprefix\url{https://link.aps.org/doi/10.1103/PhysRevB.94.174502}

\bibitem{oudah:nc2016}
Oudah M, Ikeda A, Hausmann N~J, Yonezawa S, Fukumoto T, Kobayashi S, Sato M and
  Maeno Y 2016 {\em Nat. Commun\/} {\bf 7} 13617

\bibitem{caldas:prb2012}
Caldas H and Continentino M~A 2012 {\em Phys. Rev. B\/} {\bf 86}(14) 144503
  \urlprefix\url{https://link.aps.org/doi/10.1103/PhysRevB.86.144503}

\bibitem{fulde_fflo}
Fulde P and Ferrell R~A 1964 {\em Phys. Rev.\/} {\bf 135}(3A) A550--A563
  \urlprefix\url{https://link.aps.org/doi/10.1103/PhysRev.135.A550}

\bibitem{larkin_fflo}
I L~A and Ovchinnikov Y~N 1965 {\em Sov. Phys. JETP\/} {\bf 20} 762

\bibitem{BSchaffer:prb2013}
Black-Schaffer A~M and Balatsky A~V 2013 {\em Phys. Rev. B\/} {\bf 88}(10)
  104514 \urlprefix\url{https://link.aps.org/doi/10.1103/PhysRevB.88.104514}

\bibitem{asano15}
Asano Y and Sasaki A 2015 {\em Phys. Rev. B\/} {\bf 92}(22) 224508
  \urlprefix\url{https://link.aps.org/doi/10.1103/PhysRevB.92.224508}

\bibitem{vasenko:jpcm2017}
Vasenko A~S, Golubov A~A, Silkin V~M and Chulkov E~V 2017 {\em Journal of
  Physics: Condensed Matter\/} {\bf 29} 295502
  \urlprefix\url{http://stacks.iop.org/0953-8984/29/i=29/a=295502}

\bibitem{vasenko:jetplett2017}
Vasenko A~S, Golubov A~A, Silkin V~M and Chulkov E~V 2017 {\em JETP Letters\/}
  {\bf 105} 497--501 ISSN 1090-6487
  \urlprefix\url{https://doi.org/10.1134/S0021364017080082}

\bibitem{asano11}
Asano Y, Golubov A~A, Fominov Y~V and Tanaka Y 2011 {\em Phys. Rev. Lett.\/}
  {\bf 107}(8) 087001
  \urlprefix\url{https://link.aps.org/doi/10.1103/PhysRevLett.107.087001}

\bibitem{suzuki:prb2015}
Suzuki S~I and Asano Y 2015 {\em Phys. Rev. B\/} {\bf 91}(21) 214510
  \urlprefix\url{https://link.aps.org/doi/10.1103/PhysRevB.91.214510}

\end{thebibliography}
\end{document}